\documentclass[aps,prl,showpacs,twocolumn]{revtex4}
\usepackage{graphics}
\usepackage[dvips]{color}
\usepackage{graphicx}
\usepackage{amssymb}
\usepackage{rotating}
\usepackage{epsfig}

\input epsf.sty

\newcommand{\ie}{{\em i.e.}}

\newcommand{\etal}{{\em et al.}}
\newcommand{\ud}{\mbox{d}}
\newcommand{\kBT}{\mbox{$k_{_B}T$}}
\newcommand{\Nm}{\mbox{$N\!\!-\!\!1$}}
\newcommand{\sbare}{\mbox{$\sigma_{\mbox{\tiny 0}}$}}
\newcommand{\sframe}{\mbox{$\sigma_{\mbox{\tiny f}}$}}
\newcommand{\sint}{\mbox{$\sigma_{\mbox{\tiny int}}$}}
\newcommand{\sfluc}{\mbox{$\sigma_{\mbox{\tiny fluc}}$}}
\newcommand{\ts}{\mbox{$\tilde{\sigma}$}}
\newcommand{\tss}{\mbox{$\hat{\sigma}$}}

\newcommand{\kmax}{\mbox{$k_{\mbox{\tiny max}}$}}

\newcommand{\highlight}[1]{{#1}}
\newcommand{\highlighta}[1]{{#1}}

\newcommand{\dir}{{Figs}}

\begin{document}

\title{Are stress-free membranes really 'tensionless'?}
\author{Friederike Schmid}
\affiliation{Institute of Physics, JGU Mainz, D-55099 Mainz, Germany}

\begin{abstract} 

In recent years it has been argued that the tension parameter driving the
fluctuations of fluid membranes, differs from the imposed lateral stress, the
'frame tension'. In particular, stress-free membranes were predicted to have a
residual fluctuation tension. In the present paper, this argument is
reconsidered and shown to be inherently inconsistent -- in the sense that a
linearized theory, the Monge model, is used to predict a nonlinear effect.
Furthermore, numerical simulations of one-dimensional stiff membranes are
presented which clearly demonstrate, first, that the internal 'intrinsic'
stress in membranes indeed differs from the frame tension as conjectured, but
second, that the fluctuations are nevertheless driven by the frame tension.
With this assumption, the predictions of the Monge model agree excellently with
the simulation data for stiffness and tension values spanning several orders of
magnitude.  

\end{abstract}


\maketitle

Fluid membranes are indispensable constituents of all living
things\cite{alberts}.  They are made of self-assembling amphiphilic molecules,
mostly lipids, which aggregate to bilayer sheets in an aqueous environment.
The fundamental properties of self-assembled membranes have been studied for
many decades by experimental studies of simple model membranes, by random
interface theories, and by simulations on length scales ranging from the atomic
to the mesoscopic scale. Whereas in nature and in typical experimental setups
(vesicles, supported membranes), membranes are usually under (slight) tension,
most theoretical and simulation studies have focused on the ideal 'tensionless'
case.  

In the present paper we address a basic question: What is the meaning of
'tensionless' in a membrane? Do theories and simulations/experiments really
consider the same object if they refer to tensionless membranes? This has
recently been questioned by a number of
authors\cite{imparato,fournier,stecki2}.

Experimentally and in molecular simulations, the membrane tension is identified
with the lateral stress, which is the mechanically accessible quantity. It
corresponds to the force which the membrane would exert on a surrounding
frame, hence it is also called 'frame tension' \sframe. Membranes are said
to be 'tensionless' if $\sframe =0$. In fluctuating membranes, 
the frame tension is not necessarily identical with the internal membrane 
tension \sint, \highlighta{i.e., the conjugate variable to the total
area in the free energy (at fixed number of lipids)},
since the local membrane normal fluctuates.

Theoretically, fluctuating membranes are commonly described in terms of effective
interface models: The membranes are represented by elastic sheets, subject to
an effective potential that depends on their conformation \cite{safran}. The
simplest and most famous example of such a model is the Helfrich Hamiltonian
\cite{helfrich}

\begin{equation}
 \label{eq:helfrich}
 {\cal H}_{\mbox{\tiny Helfrich}} =
   \int \ud A \: \{ \sbare + \frac{\kappa}{2} H^2 + \frac{\bar{\kappa}}{2} G
   \}
\end{equation}
for fluid membranes. Here the integral $\ud A$ runs over the surface of the
sheets, $H$ and $G$ are the total and the Gaussian curvature (i.e., the sum and
the product of the inverse principal curvature radii), and $\kappa$ and
$\bar{\kappa}$ are the bending rigidity and the Gaussian rigidity,
respectively. The first term \highlight{with the 'bare tension' \sbare\ } couples
directly to the total surface area $A$ of the membranes. For fluctuating {\em
interfaces} (e.g., between two coexisting fluid phases), this term dominates the
effective interface Hamiltonian \cite{rowlinson}. In theoretical studies 
of 'tensionless' {\em membranes}, \sbare\ is often set to zero. 
\highlight{The bare tension is related, but again not necessarily identical with
the ''fluctuation tension'' \sfluc, {\em i.e.}, the coefficient of $q^2$ in
the fluctuation spectrum of a planar membrane}\highlighta{\cite{note0}}.

The relation between the externally imposed 'frame tension' \sframe\ and the
internal tension \sint\ has intrigued scientists for quite some time. 
\highlight{In this context, it is important to note that Eq.~(\ref{eq:helfrich}) 
leaves room for different interpretations. If (\ref{eq:helfrich}) is taken to 
describe strictly incompressible membranes with fixed lipid area, but fluctuating number 
of lipids, \sbare\ is proportional to the chemical potential $\mu_0$ for adding
a lipid \cite{cai}. The number of degrees of freedom $N$ then fluctuates together 
with the total area $A$ (at fixed $A/N$). This is the point of view taken by
Cai \etal\ in a seminal work of 1994 \cite{cai}. At fixed projected area $A_p$
(but fluctuating number of lipids), they argued that the fluctuation
tension \sfluc\ renormalizes to $\sfluc=\sframe$ in the thermodynamic limit.
More recent authors~\cite{imparato,fournier,stecki2,farago1,farago2}
have used a Hamiltonian of the form (\ref{eq:helfrich}) to describe compressible 
membranes with fixed number of lipids. In that case, the number of degrees of 
freedom $N$ is constant and $\sbare$ \highlighta{is} the internal
tension\cite{farago1} \sint . This situation has first been considered by
Farago and Pincus in 2003/04, who concluded again $\sfluc=\sframe$ for 
membranes with fixed projected area \cite{farago1,farago2}.
}

This conclusion was recently questioned by Imparato \cite{imparato} and
Fournier and Barbetta \cite{fournier}.  Based on free energy calculations as
well as direct calculations of the thermally averaged lateral stress tensor
\cite{fournier} for almost planar interfaces (Monge representation, see below),
they recover an earlier result by Farago and Pincus regarding the difference
between the internal tension \sint\ and the frame tension \sframe
\cite{farago1}, but they argue that the fluctuations are driven by \sint. The
fluctuation tension is predicted to be always larger than the frame tension.
Most notably, stress-free membranes with $\sframe = 0$ are predicted to have a
residual fluctuation tension $\sfluc^{(0)}=\kBT N/A$, where $N$ is the number
of fluctuation degrees of freedom. The residual tension can be attributed to
the contribution of the fluctuations on the interfacial free energy. On large
length scales, it should dominate the fluctuation spectrum.

Experimentally, however, \highlight{the evidence for such an effect is
scarce}. In experiments, the effect of fluctuations can be measured by
monitoring the relative excess area $\langle (A-A_p)/A_p \rangle$, and the
frame tension \sframe\ can be imposed, e.g., by the Laplace pressure in
micropipette experiments \cite{evans,girard}. The experimental findings in such
setups were consistent with the assumption $\sfluc = \sframe$. \highlight{A
difference between \sfluc\ and \sframe\ was only reported in one recent study
of giant vesicles \cite{sengupta}, where \sfluc\ was inferred from the
amplitude of fluctuations and \sframe\ from the shape of the vesicles on
surfaces.}

Likewise, most simulations of fluctuating stress-free membranes \cite{goetz,
lindahl, marrink, wang, otter, brown, west} or membrane stacks \cite{loison}
could be described very satisfactorily by elastic theories for $\sfluc=0$, \ie,
the long-wavelength behavior is apparently dominated by bending modes. This holds
for simulations of atomistic \cite{lindahl, marrink} as well as coarse-grained
\cite{goetz, wang, otter,brown, west, loison} models, with two exceptions: In
simulations of membranes under varying stress, Imparato \cite{imparato} and
Stecki \cite{stecki1} independently report the fluctuation tension
\sfluc to be larger than the frame tension \sframe\ over a range of frame
tensions, in agreement with the theoretical prediction. In contrast, a more
recent coarse-grained simulation study by Neder \etal\ \cite{neder} yielded
$\sfluc = \sframe$ within the error at moderate frame tensions and $\sfluc <
\sframe$ at extreme tensions. A {\em reduction} of \sfluc\ compared to \sframe\
cannot be explained by any available theory.  On the other hand, the membrane
was also found to exhibit substantial structural changes in this high tension
regime, which presumably accounts for part of the effect.

In sum, the majority of experimental and numerical evidence so far points
towards $\sfluc = \sframe$ at low tensions, and against the existence of a
residual fluctuation tension in stress-free membranes. On the other hand,
neither the simulation data nor the experimental data were accurate enough to
unambiguously exclude that possibility. It should be noted that the elastic
parameters extracted from molecular simulation data can be severely affected by
the short-wavelength fluctuations, especially in small systems. In order to
avoid this problem, Fournier and Barbetta have carried out simulations of a
discretized fluctuating line 'membrane' in two spatial dimensions. Their
results not only clearly showed an effect \cite{fournier}, but also indicated
that the fluctuation tension \sfluc\ differs substantially from {\em both} the
internal and the frame tension, \sint\ and \sframe. The total length
of the line ({\em i.e.}, the interfacial 'area') in this study was allowed to
fluctuate freely, only controlled by the internal tension \sint.  

Historically, the overwhelming theoretical literature on 'tensionless'
fluctuating fluid membranes relies on the assumption that the fluctuation
tension is zero.  On the other hand, the physically (i.e., mechanically)
relevant tension is obviously the frame tension. If the frame tension and the
fluctuation tension are not identical, a vast body of work has to be revisited.
Therefore, a clarification of the issue is clearly desirable. The purpose of
the present work is to contribute to such a clarification. We will first
re-analyze the theory and conclude that the central result, $\sfluc \ne
\sframe$, is less rigorous than it seems. Then we will present numerical
results which strongly support the hypothesis that the frame tension and the
fluctuation tension are in fact equal for membranes with \highlight{fixed 
number of lipids and} fixed area.

The starting point of the previous studies \cite{imparato, fournier}
is the Monge representation for planar membranes \cite{safran}
\highlight{with fluctuating area and fixed number of lipids}.
The membrane is assumed to fluctuate only weakly about a plane, such that the local membrane
position can be parameterized by a function $h(x,y)$ with the constraint
$\int\!\!\!\int \ud x \: \ud y \: h(x,y) = 0$.  Furthermore, the Hamiltonian is
expanded in powers of $h$ and only quadratic terms are taken into account.
Omitting the contribution of the Gaussian curvature, which is constant for
closed membranes with fixed topology, one obtains \cite{safran}
 \begin{equation}
 \label{eq:monge}
 {\cal H}_{\mbox{\tiny Monge}} = \sbare A_p + \frac{1}{2} \int_{A_p}
 \ud x \: \ud y \{ \sbare (\nabla h)^2 + \kappa (\Delta h)^2 \},
 \end{equation}
where the integral runs of the projected area $A_p$ of the membrane. 
\highlight{Since the ensemble under consideration is one with fixed number
of lipids, \sbare\ is the internal tension \sint. Furthermore, the height 
fluctuations in this linearized model are clearly driven by \sbare, {\em i.e.}, 
$\sfluc = \sbare = \sint$.}

To fully define the Monge model, one needs to specify two additional
parameters: The in-plane and perpendicular coarse-graining lengths $\Lambda$
and $\lambda$, respectively, usually taken to be of the order of the membrane
thickness. The parameter $\Lambda$ acts as a small-wavelength cutoff and
determines the number of fluctuation degrees of freedom $N$ {\em via} $N =
A_p/\Lambda^2$.  Following Refs. \onlinecite{imparato,fournier,farago1,farago2}, 
we will assume that $N$ is proportional to the number of lipids.

For future reference, we explicitly point out the approximation
$\sqrt{1+(\nabla h)^2} \approx (1 + (\nabla h)^2/2)$
entering the expression (\ref{eq:monge}). The Monge model thus
approximates the full Hamiltonian for planar interfaces up to the order
$(\nabla h)^2 \sim (A-A_p)/A_p$. 

Throughout the remaining paper, energies will be given in units of $\kBT$.
The partition function of the Monge model can be evaluated analytically, 
giving the free energy

\noindent
\parbox{\linewidth}{
\clearpage
 \begin{eqnarray}
 \label{eq:free}
\lefteqn{
 G(N,\sint,A_p) = \sint A_p +  
\frac{(\Nm)}{2} \Big( \ln (2 \sint \lambda^2)-2
}
\\ && \qquad \qquad \quad
 + \: \big(1+\frac{\sint A_p}{4 \pi \kappa (\Nm)}\big)
    \ln\big(1+ \frac{4 \pi \kappa (\Nm)}{\sint A_p}\big)
    \Big)
\nonumber
 \end{eqnarray}
}
The only approximation entering this result was to replace the sum
$\sum_{\vec{k}} f(|\vec{k}|)$ in Fourier space $\vec{k}$ by the integral
$\frac{A_p}{2 \pi} \int_0^{k_{\mbox{\tiny max}}}\ud k \: f(k)$, where the value
$\kmax = \sqrt{4 \pi \frac{N\!-\!1}{A_p}}$ is imposed by the requirement
$\sum_{\vec{k}\ne 0} = \Nm$.

Eq.  (\ref{eq:free}) gives the interfacial free energy of a fluctuating
interface with fixed projected area $A_p$ and fluctuating total area $A$,
controlled by the internal tension \sint. Membranes are better described by an
ensemble where $A_p$ fluctuates, controlled by the frame tension \sframe, and
$A$ is fixed. In Refs.~\cite{imparato,fournier,farago1,farago2}, it is argued that
these different ensembles are equivalent and the appropriate free energy
$\tilde{G}(N,A,\sframe)$ can be obtained by a Legendre transform. Here the
number of degrees of freedom $N$ is still fixed, because the number of lipids
in the membrane is fixed.  Without having to determine $\tilde{G}$ explicitly, one can
easily calculate the relation between the frame tension and the internal
tension {\em via} $\sframe = \frac{\partial G}{\partial A_P}$, giving
 \begin{equation}
 \label{eq:frame_tension_1}
 \sframe = 
 \sint \left(
  1+\frac{1}{8 \pi \kappa}\ln(1+\frac{4 \pi \kappa (\Nm)}{\sint A_p})\right)
  - \frac{(\Nm)}{2 A_p}.
 \end{equation}
This reproduces the result in Refs.~\cite{imparato, fournier,farago1}
and demonstrates the existence of a residual internal tension,
$\sint \sim \Nm/2A_p$, at $\sframe = 0$.
Likewise, the total area is related to \sint\ and $A_p$ by
 \begin{equation}
\label{eq:area}
 A = \frac{\partial G}{\partial \sint} =
 A_p \left ( 1+\frac{1}{8 \pi \kappa} \ln(1+\frac{4 \pi \kappa (\Nm)}{\sint A_p})
 \right)
 \end{equation}
\highlight{Inverting this relation and expressing \sint\ as a function of $A$ and $A_p$},
one can rewrite the ratio $\sframe/\sint$ entirely
as a function of $(A-A_p)/A_p$, giving 
 \begin{equation}
 \label{eq:sf_monge}
 \frac{\sframe}{\sint} = 1-\frac{1}{8 \pi \kappa}(e^y-1-y)
 \approx 1 - 4\pi \kappa \big( \frac{A-A_p}{A_p} \big)^2 + \cdots
 \end{equation}
with $y=8 \pi \kappa (A-A_p)/A_p$. In the last step, we have expanded
$\sframe/\sint$ in powers of the excess area $(A-A_p)/A_p$. 

Hence the frame tension is found to differ from the internal tension to second
order in $(A-A_p)/A_p$. On the other hand, the Monge model approximates planar
interfaces only up to the first order in $(\nabla h)^2$ or $(A-A_p)/A_p$.  Thus
the results (\ref{eq:sf_monge}) and (\ref{eq:frame_tension_1}) are not rigorous!
Whatever the differences between \sframe, \sfluc, and \sint in real membranes
are, it is not possible to determine them within the Monge model. 

\highlight{Another serious issue is the equivalence of ensembles, which was
taken for granted in the above derivation. Different ensembles are only equivalent
in the thermodynamic limit of infinite membranes $A \to \infty$ at fixed $A/N$.
At zero tension, this limit does not exist within the Monge model, since tensionless 
membranes bend around on length scales larger than the persistence length and 
are no longer planar. Hence the thermodynamic limit is generally questionable for 
membranes in the floppy low-tension regime.

One can also} derive directly a Monge approximation for planar
membranes in the desired ($N,A,\sframe$) ensemble. The Hamiltonian would then read
${\cal H} = -\sframe A_p + {\cal H}_{\mbox{\tiny bending}}$, where the second
term ${\cal H}_{\mbox{\tiny bending}}$ includes the pure bending contributions,
and fluctuations are subject to the constraint that $A$ is constant. Expanding
again in powers of $h$ up to quadratic order, one obtains \highlight{\cite{note1}}
 \begin{equation}
 \label{eq:monge2}
 {\cal H}'_{\mbox{\tiny Monge}} = - \sframe A + \frac{1}{2} \int_{A_p}
 \ud x \: \ud y \{ \sframe (\nabla h)^2 + \kappa (\Delta h)^2 \}.
 \end{equation}
According to this alternative Ansatz, the fluctuations are driven by the {\em
frame} tension. This result is also only valid up to order $(A-A_p)/A_p$, like
Eq.~(\ref{eq:sf_monge}).  The Monge model can hence be used both to predict
equivalence and non-equivalence of \sfluc\ and \sframe! 

The question remains which is correct. To shed light on this issue, the present
author has carried out numerical Monte Carlo simulations of one-dimensional
'membranes' in two-dimensional space. The model is similar to that of Fournier
and Barbetta \cite{fournier}, except that the membranes had fixed interfacial
'area', {\em i.e.}, fixed contour length $L$. Compared to two-dimensional
surfaces, one-dimensional membranes have the advantage that the Helfrich
Hamiltonian, Eq.(\ref{eq:helfrich}), can be discretized in a straightforward
manner, without having to resort to sophisticated models methods like randomly
triangulated sheets \cite{gompper}.  Moreover, fluctuation effects are stronger
in two dimensions than in three dimensions, hence they can be studied more
easily.

\highlight{To set the stage, we briefly repeat the previous calculations up
to Eq.~(\ref{eq:sf_monge})} for membranes in two-dimensional space. 'Membranes'
are then simply lines with total length $L$ and projected length $L_p$.
Otherwise, the Monge model is constructed in the same way as in
Eq.~(\ref{eq:monge}), with a height function $h(x)$.  The equivalent of
Eq.~(\ref{eq:free}) reads
 \begin{eqnarray}
 \label{eq:free_1d}
\lefteqn{
 G(N,\sint,L_p) = \sint L_p + \frac{\Nm}{2} 
 \ln\Big(\frac{\lambda^2 \sint^2 L_p}{2 \pi \kappa (\Nm)}\Big)} 
\\ \nonumber && 
\qquad \qquad \qquad
 + \: \sum_{k=1}^{N/2} \ln \Big(
  \frac{(\pi k)^2}{\ts} (1+\frac{(\pi k)^2}{\ts}) \Big),
 \end{eqnarray}
where we have introduced the dimensionless rescaled tension $\ts = \sint \:
L_p^2/4 \kappa$.  We note that in two dimensions, the sum $\sum_k$ may not be
replaced by an integral, because it is dominated by the contributions at small $k$. 

The total 'area' $L$ and the frame tension \sframe\ can again be determined 
by taking the appropriate derivatives of the free energy,
$L = \partial G/\partial \sint$ and $\sframe = \partial G/\partial L_p$.
After some algebra, and approximating
$\sum_{k=1}^{N/2} (1 + (\frac{\pi k}{x})^2)^{-1}
\approx \sum_{k=1}^{\infty} (1 + (\frac{\pi k}{x})^2)^{-1}
= \frac{1}{2} (x \coth(x) - 1)$, one obtains
\begin{eqnarray}
 \label{eq:area_1d}
 L - L_p &=& 
   \frac{L_p^2}{24 \kappa} \: \frac{3}{\ts} \: (\sqrt{\ts} \coth(\sqrt{\ts}) - 1) \\
 \label{eq:tension_1d}
 \sframe - \sint &=& 
 \frac{1}{L_p} \: (\sqrt{\ts} \coth\sqrt{\ts} -1) - \frac{3 (\Nm)}{2 L_p}.
 \end{eqnarray}
which takes the value $\sframe-\sint = - 3 (\Nm)/2 L_p$ at $\ts \to 0$.
\highlight{In two dimensions, $\sframe/\sint$ as a function of
$(L-L_p)/L_p$ has an essential singularity at $(L-L_p)/L_p \to 0$. Therefore,
we cannot give a rigorous expansion as in Eq.~(\ref{eq:sf_monge}), but we can only
derive an approximate expression for the regime $\coth(\sqrt{\ts}) \approx 1$,
\begin{equation}
\label{eq:sf_monge_1d}
\frac{\sframe}{\sint} \approx
1 + 2 \frac{L-L_p}{L_p} - 24 \kappa \frac{\Nm}{L_p}
\big(\frac{L-L_p}{L_p}\big)^2 + \cdots,
\end{equation}
which shows that $\sframe/\sint$ significantly deviates from one in two dimensions 
({\em i.e.}, the deviation is linear in the small parameter $(L-L_p)/L_p$). }

For future reference, we also give the expression for the squared
amplitude of fluctuations $w^2 = \langle h^2 \rangle - \langle h \rangle^2$,
which was obtained by direct evaluation of the statistical average:
\begin{equation}
\label{eq:width_1d}
w^2 = \frac{L_p^3}{720 \kappa} \: 
   \frac{45}{\ts^2} \Big( 
  \frac{\ts}{3} + 1 - \sqrt{\ts} \coth(\sqrt{\ts}) \Big).
\end{equation}

\highlight{This completes the summary of results for the Monge model in 
two dimensions. Given the previous discussion, we would not expect them to be of
much use, because they are only valid to linear order in $(L-L_p)/L_p$ and
they have been calculated for the wrong ensemble ($(N,\sint,L_p)$ as opposed 
to $(N,L,\sframe)$). The ensembles can only be expected to be equivalent in
the thermodynamic limit $L_p \to \infty$ at fixed $N/L_p$, corresponding
to $\tilde{\sigma} \to \infty$, which is a very special limit in the
above equations. Surprisingly, it will turn out that the equations
nevertheless describe the simulation data remarkably well.
}

\begin{figure}[tb]
 \includegraphics[scale=0.3,angle=0]{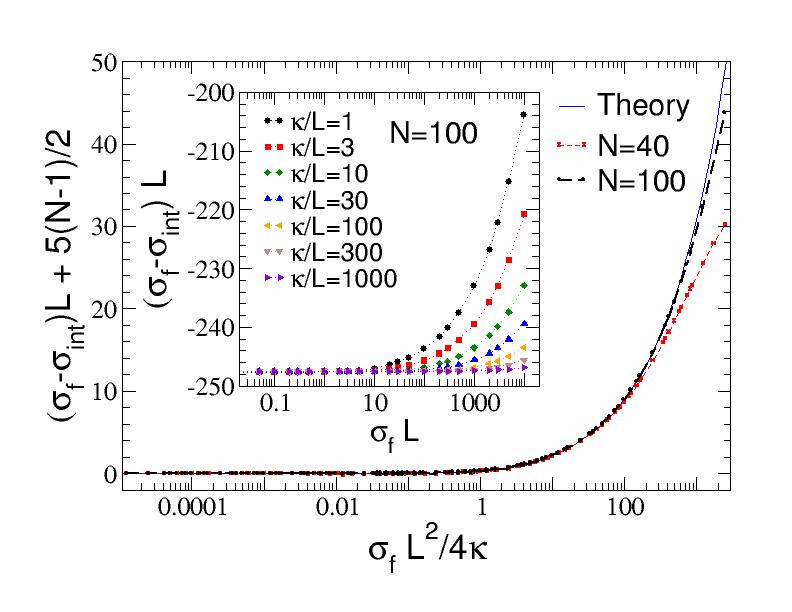}
 \vspace*{-0.3cm} \caption{\label{fig:tension}
  Difference between internal tension \sint and frame 
  tension \sframe vs. frame tension \sframe.
  Inset: Raw data for discretization $N=100$ and selected values
  of the stiffness $\kappa$.
  Main frame: Rescaled data for two discretizations $N$ and all values of 
  $\kappa$ ($\kappa/L = 1,2,3,5,10,15,20,30,50,100,200, 300,500,1000$),
  vs. rescaled tension $\tss = \sigma_f (L^2/4 \kappa)$, compared with the 
  theoretical master curve $f(x) = \sqrt{x} \coth(\sqrt{x}) - 1$ 
  from Eq.~(\protect\ref{eq:tension_1d})
  }
\end{figure}

In the numerical simulations, membranes with fixed total length $L$ were
split in $N$ segments of fixed length $L/N$, and a frame tension $\sigma_f$ was
imposed in the $x$-direction. The discretized Hamiltonian reads
\begin{equation}
{\cal H} = - \sigma_f \: L_p 
+ \kappa \: \frac{N}{L} \sum_{i=1}^N (1-\vec{e}_i \cdot \vec{e}_{i-1}),
\end{equation}
where the sum $i$ runs over the segments, $\vec{e}_i$ denotes the unit vector in
the direction of the $i$th segment, $L_p$ is the projected length $(L_p =
\frac{L}{N} \sum_i {e}_{i,x})$, and periodic boundary conditions were applied
in the $x$ direction. Monte Carlo simulations were carried out for two
discretizations, $N=40$ and $N=100$, and parameters $\kappa/L$ and $\sigma_f L$
ranging from $\kappa/L=1$ to $\kappa/L=1000$ and $\sigma_f L = 0$ and $\sigma_f
L=0.05$ to $\sigma_f L = 10000$. Typical run lengths were $0.5-1.0 \cdot 10^9$
Monte Carlo steps.  Since the results are compared with the predictions of the
Monge model, parameter combinations ($\kappa, \sigma_f$) were disregarded for
which more than 1/1000 of segments had angles larger than 45${}^0$ with the
$x$-axis. This applied to $\kappa/L \le 1$ at frame tensions $\sigma_f L \le
30$. From the simulation data, one can extract in a straightforward manner the
statistically averaged projected length, $\langle L_p \rangle$, the squared
amplitude of fluctuations, $\langle w^2 \rangle$, and the internal tension 
\highlight{$\sint = - \partial \tilde{G}/\partial L =  -
\langle \partial {\cal H}/\partial L  \rangle + 2(\Nm)/L$,
where $\tilde{G}(N,L,\sframe)$ is the free energy of the system \cite{note2}}

Fig.~\ref{fig:tension} summarizes the results for the internal tension.  The
bare data in the inset clearly show that the internal tension deviates from the
applied frame tension, as predicted by Eq.~(\ref{eq:tension_1d}). The value of the
residual tension at $\sframe \to 0$ however disagrees with the theory. Whereas
Eq.~(\ref{eq:tension_1d}) predicts $\sint L|_{\sframe = 0} \approx 3(\Nm)/2$,
the actual data rather suggest $\sint L|_{\sframe = 0} = 5(\Nm)/2$.
\highlight{The additional term $(\Nm)/2$ can be attributed to the translational 
freedom of the segments along the $x$-direction.

Next we test whether the data for $(\sframe - \sint)L_p + 3 (\Nm)/2$ can be
scaled in a way that they collapse onto a single curve for all values of
$\kappa$, $\sframe$, and $N$. Eq.~(\ref{eq:tension_1d}) suggests to try the scaling
variable $\ts = \sint L_p^2/4 \kappa$. It turns out that the data do not scale at all
with this Ansatz. The scaling is much better with the scaling variable $\sframe
L_p^2/4 \kappa$, but still not perfect (data not shown).} If one however
replaces $L_p$ by $L$ everywhere and uses the scaling variable $\tss = \sigma_f
\: L^2/4 \kappa$,  the data collapse almost perfectly onto the theoretical
master curve suggested by Eq.~(\ref{eq:tension_1d}) for all values of $\kappa$
and $\sigma_f$ and both discretizations $N$.  Only for large values of $\tss$
does one observe deviations, which can be attributed to discretization effects
since they become smaller for larger $N$ (Fig.~\ref{fig:tension}, main frame).

\begin{figure}[tb]
 \includegraphics[scale=0.3,angle=0]{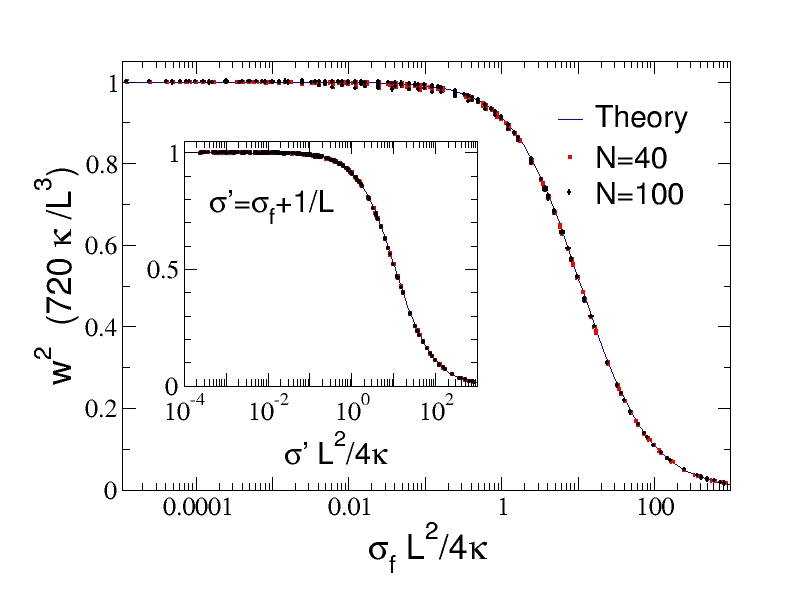} \\
 \vspace*{-0.3cm} \caption{\label{fig:width}
 Scaling plots of the squared amplitude of fluctuations $ w^2$
 vs. rescaled tension $\tss = \sigma_f (L^2/4 \kappa)$ compared
 with the theoretical master curve
  $f(x) = \frac{45}{x^2}(\frac{x}{3}+1-\sqrt{x} \coth(\sqrt{x}))$
 (Eq.~(\protect\ref{eq:width_1d})). Inset shows an alternative scaling plot 
 where $\sigma_f$ in $\tss$ is replaced by $\sigma' = \sigma_f + 1/L$.
}
\end{figure}

\highlight{These first findings already suggest that the relevant quantity
driving the fluctuations is the frame tension, not the internal tension.
Furthermore, they also suggest that the relevant {\em length} scale in the
system is the total length $L$, not the projected length $L_p$. With the
corresponding replacements, the results are in surprisingly good agreement with
the predictions of the Monge model, despite the fact that the latter have been
derived for a different ensemble with different scaling variables. We will now
proceed to investigate two quantities that probe the fluctuations directly,
{\em i.e.}, the squared amplitude of fluctuations $w^2$ and the excess length
$(L-L_p)$.  These quantities characterize the fluctuations in an integrated
way.}  

\begin{figure}[tb]
  \includegraphics[scale=0.3,angle=0]{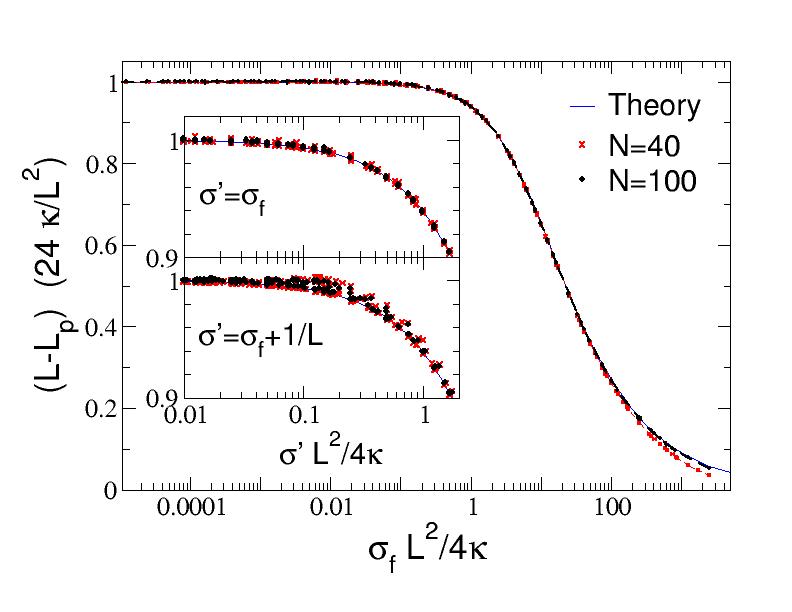} 
 \vspace*{-0.3cm} \caption{\label{fig:projection}
 Scaling plot of the excess length
 $\langle (L-L_p) \rangle$, vs.  rescaled tension 
 $\tss = \sigma_f (L^2/4 \kappa)$, compared
 with the theoretical master curve, $f(x)=\frac{3}{x} (\sqrt{x} \coth(\sqrt{x}) - 1)$ 
 (Eq.~(\protect\ref{eq:area_1d})).
 Inset compares the scaling with shifted and unshifted $\sigma'$
 in the most critical parameter region.  \\
 }
\end{figure}

Fig.~\ref{fig:width} shows scaling plots for the squared amplitude of
fluctuations. The data are rescaled as suggested by Eq.~(\ref{eq:width_1d}),
with $L_p$ replaced by $L$ and plotted against the rescaled frame tension
$\tss$. The scaling is quite good, but not perfect. It can be improved by
shifting $\sframe$ by the small amount $\delta \sigma = 1/L$ (see inset of
Fig.~\ref{fig:width}).  The data then collapse very nicely onto the
theoretical curve predicted by (\ref{eq:width_1d}). No data collapse is
obtained when plotting against the rescaled {\em internal} tension \sint\ (not
shown).  Thus the results for the squared amplitude of fluctuations indicate
that the fluctuations are driven by the frame tension \sframe, possibly
slightly shifted, and not by the internal tension \sint. 
The results for the excess length $(L-L_p)$ support this conclusion
(Fig.~(\ref{fig:projection})). In that case, the best scaling
is achieved directly with $\tss = \sigma_f (L^2/4 \kappa)$, without
additional shift. The data again collapse excellently onto the theoretical
master curve suggested by Eq.~(\ref{eq:area_1d}).

In sum, the simulation data show clearly that the membrane fluctuations are
driven by the frame tension \sframe. 
These results raise the question why Barbetta and Fournier in
Ref.~\cite{fournier} obtained a different result from simulations of
essentially the same model. The answer is that these simulations were carried
out in a different ensemble where both the contour length and the
projected length were allowed to fluctuate, {\em i.e.}, the $(N,\sint,\sframe)$
ensemble as opposed to the $(N,L,\sframe)$ ensemble considered in the present
work. \highlight{As pointed out earlier, the ensembles are not equivalent}.
To further analyze the issue, additional simulations were carried out for
membranes in the $(N,L,L_p)$ ensemble and in the $(N,\sint,L_p)$ ensemble. The
purpose of these simulations was to find out whether the fluctuations in other
ensembles might be driven by $\sint$ instead of $\sframe$.

\begin{figure}[tb]
 \includegraphics[scale=0.3,angle=0]{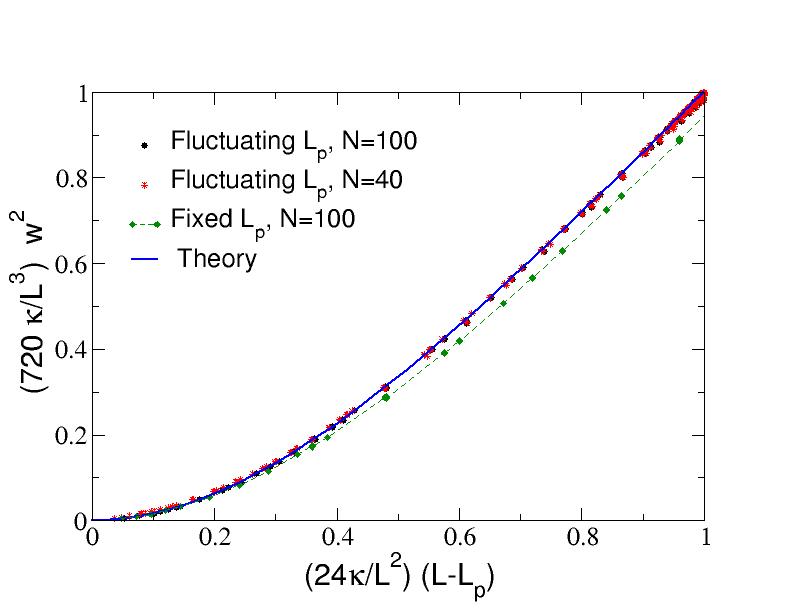} 
 \vspace*{-0.3cm} \caption{\label{fig:width_vs_projection}
  Rescaled squared amplitude of fluctuations vs. excess length.  Circles and
stars correspond to data for fluctuating projected length $L_p$ and
discretizations $N=40$ and $N=100$, respectively.  Diamonds correspond to data
from simulations with fixed projected length $L_p$. The solid line shows the
theoretical prediction from Eq.~(\protect\ref{eq:area_1d}) combined with
Eq.~(\protect\ref{eq:width_1d}).  The dashed line is a guide for the eye.
  }
\end{figure}

Fig.~\ref{fig:width_vs_projection} shows the squared amplitude of
fluctuations $w^2$ vs. the excess length $(L-L_p)$, in a scaling plot
\highlight{which is independent of the actual choice of the rescaled
tension}.  The data are taken from simulations in the $(N,L,\sframe)$ ensemble
and in the $(N,L,L_p)$ ensemble. In the $(N,L,\sframe)$ ensemble, they collapse
onto a single curve which agrees nicely with the theoretical prediction from
Eqs.~(\ref{eq:area_1d}) combined with (\ref{eq:width_1d}).  The data from the
$(N,L,L_p)$ ensemble also collapse, but the scaling curve is different and not
consistent with Eqs.~(\ref{eq:area_1d}) and (\ref{eq:width_1d}) \cite{note3}.

If $L$ is allowed to fluctuate, the situation is further complicated by the
fact that in one dimension, the total bending energy of bubbles decreases with
increasing bubble size. As a consequence, the simulation data in the
$(N,\sint,L_p)$ ensemble feature a transition between a flat state and an
inflated bubble state at bending stiffness $\kappa/L \sim 3$ and low tensions
(data not shown). The simulations of Fournier and Barbetta were carried
out at $\kappa/L=2.5$ (in yet another ensemble), hence the vicinity of a similar
transition possibly accounts for some of the intriguing phenomena reported in
Ref.~\onlinecite{fournier}. 

In the introduction, we have raised the question whether theories and
simulations/experiments really consider the same object if they refer to
'tensionless' membranes? Based on the simulation results presented above, we
conclude that the answer is most likely ''yes''. At least in one dimensional
membranes with fixed 'area', the fluctuations are driven by the imposed lateral
stress, and the behavior of stress-free (tensionless) membranes is consistent
with that of the Monge model at tension zero. These results will presumably
also hold for two-dimensional membranes. We have presented an argument in
Eq.~(\ref{eq:monge2}) that rationalizes the prediction $\sfluc=\sframe$ in the
physically relevant $(N,A,\sframe)$ ensemble for arbitrary dimensions.
\highlight{We should note, however, that our numerical simulations indicate
that the relevant length scale is the total 'area' rather than the projected
'area'. This presumably renormalizes the actual '$q^2$ coefficient', $\sfluc =
\sframe \: (L_p/L)^2$ in two dimensions \cite{note4}, or $\sfluc = \sframe \:
(A_p/A)$ in $d$ dimensions.  A detailed analysis of the full
fluctuation spectra as a function of the wavevector $q$ will be presented
elsewhere. }

The present author hopes that this paper will stimulate discussions and further
work on these intriguing and important issues. \highlight{For example, it
is not clear why the Monge model provides such a good prediction of the
scaling {\em functions} for the various quantities, even though it is 
derived for the wrong ensemble with different scaling {\em variables}. 
}
Furthermore, mesoscopic simulations of two-dimensional fluctuating fluid
sheets, {\em e.g.}, using randomly triangulated surfaces \cite{gompper} or
Fourier techniques \cite{troester}, would clearly be desirable.

The author thanks J. Neder, J.-B. Fournier and M. Weikl for inspiring
discussions. Partial support from the German Science Foundation (CRC 625) is
gratefully acknowledged.


\begin{thebibliography}{66}
\bibitem{note2}
\highlight{This is analogous to the virial equation for the pressure in
simple $(NVT)$ systems. To derive it one rewrites the partition function
${\cal Z} = \prod_{i=2}^{N} [\int \ud \vec{r}_i \delta(|\vec{r}_i - \vec{r}_{i-1}|-L/N)]
e^{- {\cal H}}$ with the joint positions $\vec{r}_i$ (fixing $\vec{r}_0=0$) in terms of
rescaled variable $\vec{\xi}_i = \vec{r}_i \: L/N$ and then calculates
$- \partial \tilde{G}/\partial L = \partial(\ln {\cal Z})/\partial L$.}
\bibitem{note3}
The agreement with the theory does not become better if one rescales $w^2$ and
$(L-L_p)$ with $L_p$ instead of $L$.
\bibitem{note4}
\highlight{For a given wave number $n$, the height fluctuations are
$\langle h_n h_{-n} \rangle \propto 1/(\tau \: n^2 + {\cal O}(n^4))$
with $\tau \: n^2 = \sframe (2 \pi n/L)^2 = \sfluc (2 \pi n/L_p)^2$.
}
\end{thebibliography}
\end{document}